\documentclass[prl,10pt,twocolumn,amsmath,amssymb,superscriptaddress]{revtex4-1}
\usepackage{graphicx}% Include figure files
\usepackage{dcolumn}% Alig
\usepackage{bm}% bold math
\usepackage{amsfonts}
\usepackage[latin1]{inputenc}
\usepackage{siunitx}
\usepackage{graphicx}
\usepackage{pdfpages}
\usepackage{amssymb}
\usepackage{amsmath}
\usepackage{braket}

\begin{document}

\author{Robert~B\"ucker}
 \affiliation{Vienna Center for Quantum Science and Technology, Atominstitut, TU Wien, 1020 Vienna, Austria} 
 \author{Julian~Grond}
  \altaffiliation{Current Affiliation: Theoretische Chemie, Universit\"at Heidelberg, Im Neuenheimer Feld 229, 69120 Heidelberg, Germany}
 \affiliation{Vienna Center for Quantum Science and Technology, Atominstitut, TU Wien, 1020 Vienna, Austria} 
 \affiliation{Institut f\"ur Physik, Karl-Franzens-Universit\"at Graz, 8010 Graz, Austria}
 \affiliation{Wolfgang Pauli Institute, 1090 Vienna, Austria} 
 \author{Stephanie~Manz}
 \altaffiliation{Current Affiliation: Center for Free-Electron Laser Science, Notkestrasse 85, 22607 Hamburg, Germany, and MPSD at the University of Hamburg, Notkestrasse 85, 22607 Hamburg, Germany}
 \affiliation{Vienna Center for Quantum Science and Technology, Atominstitut, TU Wien, 1020 Vienna, Austria} 
 \author{Tarik~Berrada}
 \affiliation{Vienna Center for Quantum Science and Technology, Atominstitut, TU Wien, 1020 Vienna, Austria} 
\author{Thomas~Betz}
\altaffiliation{Current Affiliation: Center for Free-Electron Laser Science, Notkestrasse 85, 22607 Hamburg, Germany, and Max-Planck-Institut f\"ur Kernphysik, Saupfercheckweg 1, 69117 Heidelberg, Germany}
 \affiliation{Vienna Center for Quantum Science and Technology, Atominstitut, TU Wien, 1020 Vienna, Austria} 
\author{Christian~Koller}
 \affiliation{Vienna Center for Quantum Science and Technology, Atominstitut, TU Wien, 1020 Vienna, Austria} 
 \author{Ulrich~Hohenester}
 \affiliation{Institut f\"ur Physik, Karl-Franzens-Universit\"at Graz, 8010 Graz, Austria}
\author{Thorsten~Schumm}
 \affiliation{Vienna Center for Quantum Science and Technology, Atominstitut, TU Wien, 1020 Vienna, Austria} 
 \affiliation{Wolfgang Pauli Institute, 1090 Vienna, Austria} 
  \author{Aur\' elien~Perrin}
  \altaffiliation{Current Affiliation: Laboratoire de physique des lasers, CNRS, Universit\' e Paris 13, 99 avenue J.-B. Cl\' ement, 93430 Villetaneuse, France}
 \affiliation{Vienna Center for Quantum Science and Technology, Atominstitut, TU Wien, 1020 Vienna, Austria} 
   \affiliation{Wolfgang Pauli Institute, 1090 Vienna, Austria}
\author{J\"org~Schmiedmayer}
 \affiliation{Vienna Center for Quantum Science and Technology, Atominstitut, TU Wien, 1020 Vienna, Austria} 
\title{Twin-atom beams}

\date{\today}

\maketitle

\textbf{
In recent years, tremendous progress has been made in exploring and exploiting the analogy of classical light and matter waves for fundamental investigations and applications~\cite{Cronin2009}. 
Extending this analogy to \emph{quantum} matter wave optics is promoted by the intrinsic non-linearity of interacting particles, a stepping stone towards non-classical states~\cite{deng:99,Orzel2001}. 
In light optics, twin-photon beams~\cite{Heidmann1987} are a key element in providing the non-local correlations and entanglement required for applications like precision metrology and quantum communication~\cite{Reid2009}. 
Similar sources for massive particles have so far been limited by the multi-mode character of the involved processes or a predominant background signal~\cite{Vogels2002,Gemelke2005,Campbell2006,Spielman2006,perrin:07,Dall2009,Jaskula2010a,Klempt2010}. 
Here we present highly efficient emission of twin-atom beams into a single transversal mode of a waveguide potential.
The source is a one-dimensional degenerate Bose gas~\cite{Petrov2000} in the first radially excited state.
We directly measure a suppression of fluctuations in the atom number difference between the beams to 0.37(3) with respect to the classical expectation, equivalent to 0.11(2) after correcting for detection noise.
Our results underline the high potential of ultracold atomic gases as sources for quantum matter wave optics and will enable the implementation of schemes previously unattainable with massive particles~\cite{Reid2009,Dunningham2002,Horne1989,Hong1987,Rarity1990,Gneiting2008}. 
}

Binary collisions between atoms provide a natural means to generate dual number states of intrinsically correlated atoms~\cite{Dunningham2002}.
Experimental schemes include spontaneous emission of atom pairs by collisional deexcitation~\cite{Spielman2006} or four-wave mixing~\cite{perrin:07,Dall2009,Jaskula2010a}.
Stimulated emission into twin-modes has been demonstrated in seeded four-wave mixing~\cite{deng:99,Vogels2002}, and parametric amplification in optical lattices~\cite{Gemelke2005,Campbell2006} or spinor condensates~\cite{Klempt2010,*Lucke2011,Bookjans2011,*Hamley2012,Gross2011}.
For motional states, suppression of relative number fluctuations could so far only be demonstrated for multi-mode twin-atoms~\cite{Jaskula2010a}.
A different route to non-classical states is provided by ensembles in multi-well potentials, that become number-squeezed during their time evolution~\cite{Esteve2008,Maussang2010}.

Here, we demonstrate how collisional deexcitation of a one-dimensional degenerate Bose gas can be used to efficiently create matter wave beams of twin-atoms.
The restricted geometry of a waveguide potential enforces emission of the beams into a single transversal mode, in analogy to an optical parametric amplifier~\cite{Heidmann1987}.
We prepare the initial population inversion to a radially excited state by shaking the trap following an optimal control strategy.
Time-of-flight fluorescence imaging is used to directly observe the suppressed relative number fluctuations in the emitted beams.

\begin{figure}
\includegraphics{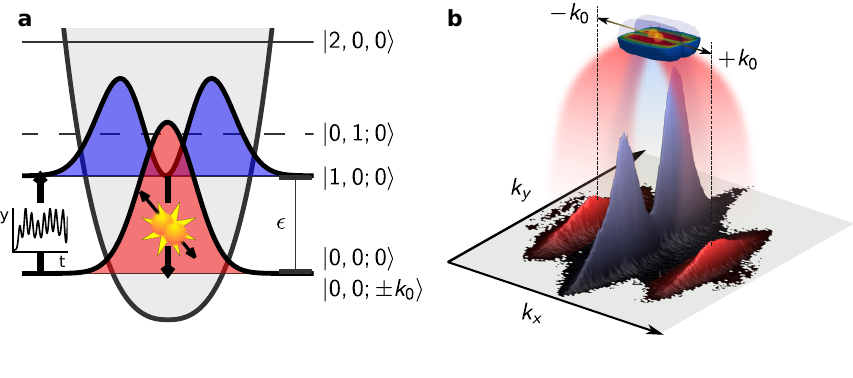}
\caption{Schematic of the excitation and emission process. (a) The quasi-BEC is transferred from the ground state $\ket{0,0;0}$ into $\ket{1,0;0}$, the first excited state of the trapping potential along the radial $y$-direction. This is accomplished by means of fast non-adiabatic movement of the potential minimum along an optimized trajectory (inset). The excited state decays by emission of twin-atoms into the radial ground state modes $\ket{0,0;\pm k_0}$. (b) After excitation and pair emission, the cloud is released from the trapping potential and imaged after expansion. The central part of the system clearly shows the spatial structure of the radially excited state (blue). Two clouds containing the twin-atoms (red) are emitted.}
\label{fig:overview}
\end{figure}

The starting point of our investigations is a dilute, quantum degenerate gas of neutral $^{87}$Rubidium atoms magnetically trapped in a tight waveguide potential with a shallow axial harmonic confinement ($\nu_x=\SI{16.3}{Hz}$) on an atom chip~\cite{Folman2002}. 
Our scheme relies on an effective two-level system in the radial vibrational eigenstates of the waveguide. 
This is accomplished by creating unequal level spacings in the radial $y,z$-plane by radio frequency dressing~\cite{Schumm2005b}, which introduces anharmonicity and anisotropy.
The resulting single-particle first and second excited state energies are  $E_{y,z}^{(1)}=h \cdot [1.83, 2.58]~\si{kHz}$ and $E_{y,z}^{(2)}=h \cdot [3.82, 5.22]~\si{kHz}$.
Due to the increasing level spacings, the ground state $\ket{n_y,n_z;k_x}=\ket{0,0;0}$ ($n_{y,z}$ and $k_x$ denoting the radial quantum numbers and the axial momentum, respectively) and the first excited state along $y$, $\ket{1,0;0}$ have the lowest energy difference among all possible combinations (figure~\ref{fig:overview}a), establishing a closed two-level system.

Using standard techniques we generate a Bose gas of typically 700 atoms at a temperature $T\lesssim\SI{40}{nK}\approx h/k_B \cdot \SI{830}{Hz}$ (obtained independently from fits to a non-excited degenerate gas~\cite{Stimming2010} and its residual thermal fraction~\cite{Perrin2010}). 
The thermal occupation of state $\ket{1,0;0}$ is negligible, and the chemical potential, quantifying the mean-field interaction is $\mu\sim h\cdot \SI{500}{Hz}\ll E_y^{(1)}$.
Our Rb sample is therefore a one-dimensional, weakly interacting quasi-Bose-Einstein condensate (quasi-BEC, interaction parameter $\gamma\sim 0.008$~\cite{Petrov2000}), which can be seen as a locally coherent matter wave~\cite{Hofferberth2007b} with a coherence length approximately one order of magnitude below the system size.

\begin{figure}
\includegraphics{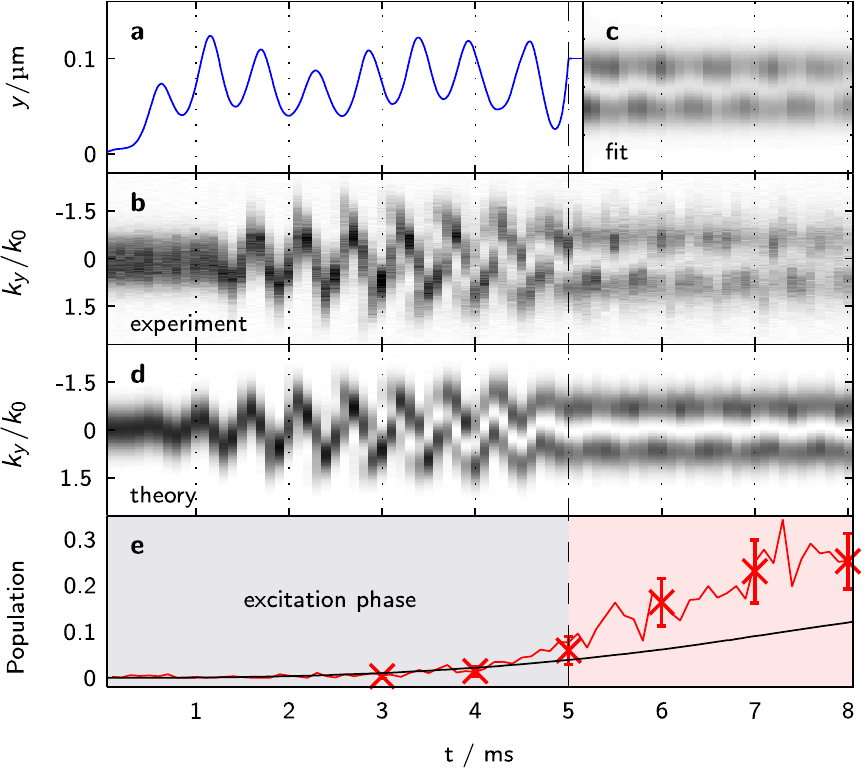}

\caption{Dynamics of the excitation and emission process: comparison between theory and experiment. (a) Optimized control trajectory for the trap movement along $y$. (b) Measured momentum distribution. Fluorescence images integrated along $k_x$ over a region encompassing the central cloud. Average over 5 experimental runs. (c) Fit of the time-dependent momentum distribution, scaled to the experimental total density at each time step. (d) Calculated momentum distribution, using a one-dimensional GPE model. As the model does not take into account the pair emission process, agreement to the experiment is expected approximately up to the end of the excitation pulse. (e) Red line: Population of the emitted clouds obtained from the same data set as (b). Red crosses: population of the emitted clouds obtained from a separate measurement with 100 experimental runs each (error bars are the ensemble standard deviation). Black line: theoretical estimation for spontaneous processes only.}

\label{fig:dynamics}
\end{figure}

Having prepared the gas, we create a population inversion by transferring the quasi-BEC almost entirely to state $\ket{1,0;0}$ (see figure~\ref{fig:overview}a).
The transition is driven by shaking the trap along the radial $y$-direction on the scale of the ground state size ($\sim\SI{100}{nm}$).
The trajectory (total duration \SI{5}{ms}, see figure \ref{fig:dynamics}a) has been optimized employing an iterative optimal control algorithm (see methods). 
In the experiment, the displacement is achieved by driving a current in an auxiliary chip wire, parallel to the main trapping wire.

We monitor the radial momentum distribution of the quasi-BEC by releasing the cloud from the trapping potential at different times $t$ during and after the excitation pulse. 
Images are taken after \SI{46}{ms} of ballistic expansion (figure~\ref{fig:dynamics}b), using a single-atom-sensitive fluorescence imaging system~\cite{Buecker2009}.
After the excitation, we observe a small residual beating between the macroscopically occupied $\ket{1,0;0}$ state and a remaining non-excited population in $\ket{0,0;0}$.
From a fit to the beating pattern (figure~\ref{fig:dynamics}c) we estimate an efficiency of the coherent transfer of $\eta_e\approx 97\,\%$ and deduce the energy difference $\epsilon=h\cdot\SI{1.78}{kHz} $ between $\ket{0,0;0}$ and $\ket{1,0;0}$ (see methods).
A more detailed, quantitative model of the excitation dynamics is given in~\cite{Buecker2012}.
The slight deviation between $E^{(1)}_y$ and $\epsilon$ is explained by particle interactions.
A calculation based on the one-dimensional Gross-Pitaevskii-Equation (GPE)  (figure~\ref{fig:dynamics}d) shows excellent agreement to the observed dynamics.

\begin{figure}
\includegraphics{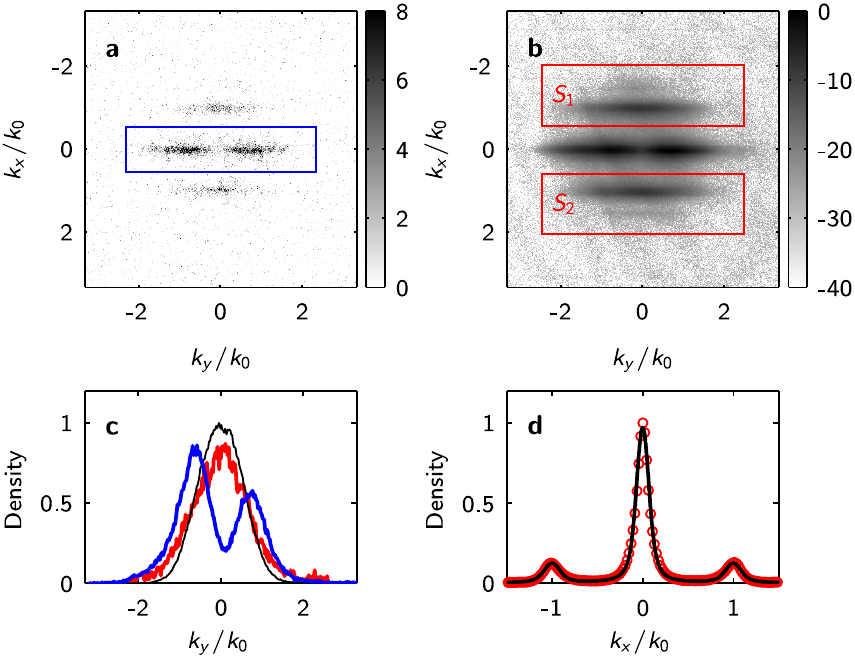}

\caption{Atom cloud image analysis. (a) Typical experimental image of $\sim 700$ atoms released from the trap \SI{7}{ms} after starting the excitation sequence. The cloud is allowed to expand for~\SI{46}{ms}, making the initial momentum distribution accessible. The quasi-BEC in the excited state $\ket{1,0;0}$ is clearly distinct from the emitted clouds at momenta $\pm \hbar k_0$. Units are photons per pixel. The blue box indicates the integration range for the data shown in figure \ref{fig:dynamics}b. (b) Average over $\approx 1500$ images similar to (a). The colour scale is logarithmic (dB referenced to peak density). The regions used for correlation analysis are indicated as red boxes. (c) Normalized, radial momentum distributions of the central (blue) and emitted (red) clouds. Average of 50 images of clouds released at $t=\SI{6}{ms}$. As comparison, the distribution of a non-excited cloud is shown (black, average over 100 images). (d) Normalized profile of (b) along $k_x$ (red dots) and three-peak fit (black line) based on stochastic simulations~\cite{Stimming2010}.}

\label{fig:images}
\end{figure}

The population inversion to $\ket{1,0;0}$ represents a highly non-equilibrium state of the system, analogous to a laser gain medium after a pump pulse.
For the ensuing relaxation, the only allowed channel is a two-particle collisional process, emitting atom pairs with opposite momenta. 
In contrast to experiments with free-space collisions~\cite{perrin:07,Jaskula2010a} or two-dimensional gases~\cite{Spielman2006}, the constricted geometry and non-degenerate level scheme of our source restricts the outgoing matter waves to the radial ground state of the waveguide, yielding twin-atom beams in a single transversal mode.
Within a binary collision, two atoms are scattered from $\ket{1,0;0} \ket{1,0;0}$ to $\ket{0,0;+k_0} \ket{0,0;-k_0}$, where energy conservation requires the final momenta to be centred around $\pm k_0=\pm \sqrt{2m \epsilon}/\hbar$.
The emission process can be understood as a matter wave analogue to a degenerate optical parametric amplifier, where the initially empty twin-modes are seeded by vacuum fluctuations~\cite{Heidmann1987} and gain an exponentially growing population if phase matching conditions are fulfilled.
Due to finite size, the axial multi-mode character of the quasi-BEC source~\cite{Petrov2000}, and the depletion of the initial state, a quantitative description of the process is challenging. 
In ref.~\cite{Buecker2012}, a model based on a density matrix expansion has been developped, which mostly covers those effects.
A comparison of the observed rates to a simple calculation using Fermi's golden rule (see methods) is shown in figure~\ref{fig:dynamics}e and demonstrates the insufficiency of a purely spontaneous model (in contrast to the findings of~\cite{Spielman2006} for a transversal multi-mode system).

Once the trap potential is switched off and the atoms propagate freely (figure~\ref{fig:overview}b), the twin-beam modes can be detected essentially background-free in fluorescence images (figure~\ref{fig:images}a,b) as they separate from the source.
In figure~\ref{fig:images}c, the radial momentum distribution of the twin-beams is compared to an independent measurement of the initial $\ket{0,0;0}$ cloud.
The small deviation is attributed to a slight overlap of the central cloud into the integration regions of the emitted clouds (red boxes in figure~\ref{fig:images}b) and interactions with the the mean field of the quasi-BEC in the $\ket{1,0;0}$ state.
Furthermore, an excited cloud at $t=\SI{6}{ms}$ is shown.
From the axial position of the side peaks (figure \ref{fig:images}d), we can deduce an emitted atom momentum of $k_0=2\pi\cdot 0.883(3)~\si{\micro m^{-1}}$, equivalent to $\epsilon=h \cdot1.78(1)~\si{kHz}$, in perfect agreement with the value determined from the beating fit (figure \ref{fig:dynamics}c). 
The width of the emitted clouds (figure~\ref{fig:images}d) is increased by a factor of $\approx 1.4$ with respect to the source cloud.
At momenta corresponding to $\epsilon'\approx h\cdot\SI{3.9}{kHz}$, very weakly populated additional atom clouds ($\lesssim 1$ atom per image) are observed on an averaged picture (figure \ref{fig:images}b).
They imply a transient population of $\ket{2,0;0}$ during the control sequence, which directly decays into the radial ground state. 
In the following analysis, they are merged with the atoms originating from $\ket{1,0;0}$.

\begin{figure}
\includegraphics{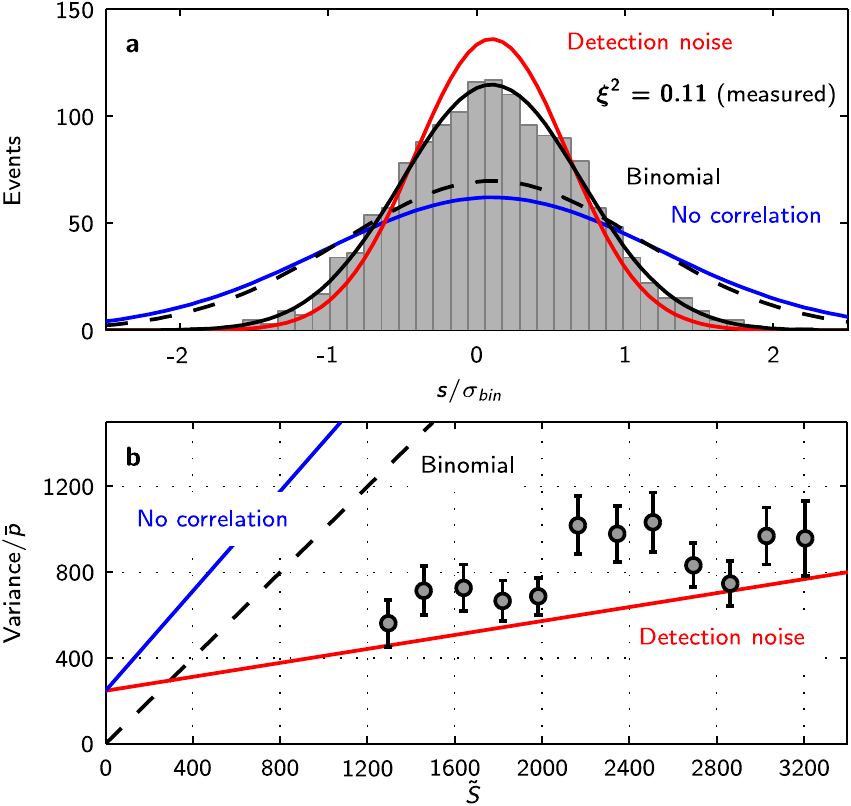}

\caption{Correlation analysis. (a) Histogram of observed signal imbalances $s$ between the emitted clouds, in units of the binomial standard deviation $\sigma_{bin}=(\bar{p}\tilde{S})^{1/2}$. 
The curves indicate normal distributions corresponding to the experimental result of $\xi^2=0.11(2)$ (black, solid), the limits of perfect correlation, where only detection noise remains (red, solid), of uncorrelated signals, defining the reference point for $\xi^2$ (blue, solid), and a binomial distribution for $\bar{p}\tilde{S}$ trials (black, dashed).
(b) Observed signal imbalance variances for data bins corresponding to different total signal in the emitted clouds $\tilde{S}$. Error bars are the standard error. The lines correspond to those in panel (a). The corrected variances are given by the vertical distances between the data points and the detection noise.}

\label{fig:histogram}
\end{figure}

The non-classical correlation in the emitted twin-atom beams is revealed by a sub-binomial distribution of the number imbalance $n=N_1-N_2$ between atoms detected at $\pm k_0$.
The variance of $n$ can be expressed as $\sigma^2_n=\xi^2\bar{N}$, where $\bar{N}$ denotes the mean total atom number in the emitted clouds.
The noise reduction factor $\xi^2$ quantifies the suppression of $\sigma_n^2$ with respect to a binomial distribution, and thus the amount of correlation between the populations $N_1$ and $N_2$.

In the fluorescence images, we count photons in regions encompassing the emitted clouds, which have been released at $t=\SI{7}{ms}$. 
For given atom numbers $N_{1,2}$, the expectation values for the photon numbers are $S_{1,2}=\bar{p} N_{1,2}+\bar{b}/2$, where $\bar{p}=12.3(9)$ denotes the average number of photons per atom and $\bar{b}/2$ accounts for background events. 
Our main observable is the variance $\sigma_s^2$ of the signal imbalance $s=S_1-S_2$.
Its expectation value for a binomial distribution of atoms is given by $\sigma_{bin}^2=\bar{p}\tilde{S}$, where $\tilde{S}=\bar{S_1}+\bar{S_2}-\bar{b}$.
From the experimental data as shown in figure~\ref{fig:histogram}a we obtain an uncorrected reduction factor $\sigma_s^2/\sigma_{bin}^2=0.37(3)$. 
However, a significant contribution to $\sigma_s^2$ does not originate from the atom number fluctuations, but from the detection process itself.
For fluorescence imaging as employed in our experiment, this contribution can directly be calculated from photon shot noise and detection background (see methods).
It is accounted for by subtracting a correction $\sigma_d^2$ from $\sigma_s^2$.
From the corrected variances, we infer a reduction factor of $\xi^2=(\sigma_s^2-\sigma_d^2)/\sigma_{bin}^2=0.11(2)$, which is the main result of this paper. 
It is equivalent to an intensity squeezing in the sense of~\cite{Heidmann1987}.
In strong contrast to the suppressed \emph{relative} fluctuations, applying an analogous calculation on the variance $\sigma_S^2$ of the \emph{summed} signal in the emitted clouds $S=S_1+S_2$ (binned into groups with similar total atom number) yields super-Poissonian fluctuations $(\sigma_S^2-\sigma_d^2) / \bar{p}\tilde{S}\sim 7$, again highlighting the presence of bosonic amplification.

To study the correlation data in more detail, we bin the experimental shots according to the emitted atom signal $S$ and calculate the variances $\sigma_s^2$ and mean signals $\tilde{S}$ for each bin (consisting of typically 100 runs) separately, as shown in figure \ref{fig:histogram}b.
The differences between the data points and the detection noise $\sigma_d^2$ represent the corrected variances as introduced above.
They appear to be independent of $\tilde{S}$, which is supported by $\chi^2$-test results on various plausible models.
As any uncorrelated emissions should scale with $\tilde{S}$, this suggests that the non-zero value of $\xi^2$ can be explained by a slight additional background signal, e.g. due to the residual overlap of the excited quasi-BEC and the emitted clouds.

%In conclusion, we have demonstrated efficient generation of twin-atom beams in a single transverse mode with non-classical density correlations, in analogy to an optical parametric amplifier~\cite{Heidmann1987}. 
%As a source we use a radially excited one-dimensional quasi-BEC that acts as a non-linear gain medium and can only decay via emission of two particles with exactly opposite momentum. 
%The excitation is achieved by means of non-adiabatic spatial displacement of the system, optimized using optimal control theory. 
%This scheme can be applied to any sufficiently controllable quantum system of interacting bosons.

The availability of single-mode twin-atom beams adds an essential building block for quantum matter wave optics. 
As our scheme does not rely on the internal structure of the atoms, it can be applied to any sufficiently controllable system of interacting bosons.
Possible applications include interferometry with dual-Fock states~\cite{Dunningham2002}, Hong-Ou-Mandel type experiments~\cite{Hong1987} or continuous-variable entanglement~\cite{Gneiting2008,Reid2009}.
Inclusion of internal (e.g. hyperfine states with appropriate scattering properties) or few-mode external (e.g. two-mode double well states) degrees of freedom seems a viable strategy to generate non-local entangled states of massive particles for Bell-type measurements~\cite{Pu2000,Duan2000a,Horne1989,Rarity1990,Gneiting2008}.

\section*{Acknowledgements}

We acknowledge support from the FWF projects P21080-N16 and I607, the EU projects AQUTE, QuDeGPM and Marie Curie (FP7 GA no. 236702), the FWF doctoral programme CoQuS (W 1210) and the FunMat and NAWI GASS research alliances. We wish to thank E. Altman, A. Gottlieb, B. Hessmo, K. Kheruntsyan, I. Mazets, M. Oberthaler, H. Ritsch and G. von Winckel for stimulating discussions.

\section*{Author information}

Correspondence and requests for materials should be addressed to J.S. (schmiedmayer@atomchip.org).

\section*{Author contributions}

R.B., S.M. and T. Berrada collected the data presented in this letter. 
J.G. and U.H. provided the OCT calculations for the excitation sequence.
All authors contributed to analysis and interpretation of the data and helped in editing the manuscript.

\section*{Competing financial interests}
The authors declare that they have no competing financial interests.

{\small

\section{Methods}

\subsection{Preparation of the trapping potential}

\begin{figure}

\centering
\includegraphics[width=90mm]{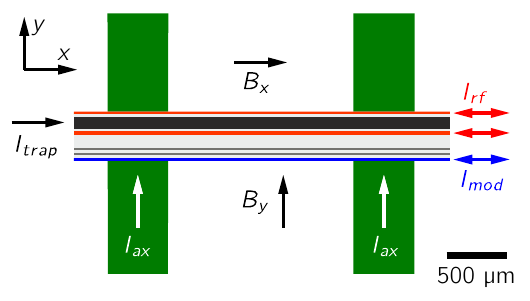}

\caption{Schematic of the atom chip layout.  The waveguide potential is formed by the current $I_{trap}$ through the main trapping wire (black) and a static magnetic field $B_{y}$. On a separate chip layer, currents $I_{ax}$ in broad wires (green) provide axial confinement. An external field $B_{x}$ completes the Ioffe-Pritchard configuration. The radio frequency dressing currents $I_{rf}$ are applied to wires (red) in parallel to the trapping wire. Finally, the modulation of the trap position is accomplished by a current $I_{mod}$ in an auxiliary wire (blue).}
\label{fig:wires}

\end{figure}

Both the optimized excitation scheme needed for transferring the quasi-BEC into $\ket{1,0;0}$ (see next section), and restricting the emission to a single transverse mode $\ket{0,0;\pm k_0}$ require a sufficiently anharmonic trapping potential along $y$ with increasing level spacings.
Hence, the initially radially symmetric Ioffe-Pritchard field configuration created by our chip wire configuration (see \cite{Trinker2008a} and figure~\ref{fig:wires}) is being modified by radio frequency dressing \cite{Lesanovsky2006,Schumm2005b}.
Typically used for creating double well potentials, this technique also allows for the introduction of anharmonicity and anisotropy to a single trap when the dressing strength is kept slightly below the point where actual splitting of the potential occurs.
We apply an ac current of $I_{rf}=\SI{23}{mA}$ peak-to-peak amplitude at a detuning of $\delta=-\SI{54}{kHz}$ with respect to the atomic Larmor frequency near the trap minimum ($\nu_0=\SI{824}{kHz}$) to two wires running on each side of the main trapping wire at a distance of \SI{55}{\micro m}.

The resulting potential can be calculated numerically by means of a Floquet analysis \cite{Shirley1965}.
In the two radial directions it can be approximated by quartic polynomials of the form $E=p_4 r^4 + p_2 r^2$: 
In the $y$-direction, along which the excitation is performed the coefficients are $p_4=h \cdot\SI{13.1}{Hz/r_0^4}$ and $p_2=h \cdot\SI{343}{Hz/r_0^2}$.
In the $z$-direction perpendicular to the excitation motion the coefficients are $p_4=h \cdot\SI{10.4}{Hz/r_0^4}$ and $p_2=h \cdot\SI{793}{Hz/r_0^2}$.
Here, $r_0=\SI{172}{nm}$ is the mean radial ground state radius as calculated by the Gross-Pitaevskii equation.

% $p_4=2\pi\hbar\cdot\SI{15.0}{\kHz/\micro m^4}$ and $p_2=2\pi\hbar\cdot\SI{11.6}{\kHz/\micro m^2}$.
% $p_4=2\pi\hbar\cdot\SI{11.8}{\kHz/\micro m^4}$ and $p_2=2\pi\hbar\cdot\SI{26.8}{\kHz/\micro m^2}$

Along the axial $x$-axis, the trap frequency given in the text is determined by observation of a deliberately excited sloshing mode of the quasi-BEC.

\subsection{Optimized excitation of the condensate}

To transfer the cloud into the radially excited state $\ket{1,0;0}$ we displace the radial trap minimum along an optimized trajectory.

This movement is achieved by applying a current of typically less than \SI{10}{mA} to a wire parallel to the trapping wire at \SI{140}{\micro\metre} distance (see figure~\ref{fig:wires}). 
The resulting magnetic field is mostly oriented along the $z$-direction at the cloud position, moving the potential predominantly along $y$ \cite{Folman2002}. 
The calculated offset is \SI{26}{nm/mA} along $y$ and \SI{9}{nm/mA} along $z$, where the contribution along $z$ does not significantly distort the excitation process (see below). 

The most efficient way of moving the trap for excited state preparation is obtained from numerical calculations which employ \emph{optimal control} \cite{peirce:88} of the Gross-Pitaevskii equation. 
Optimal control theory is a powerful tool which allows to minimize a given cost functional with the constraint that the system is governed by the corresponding equations of motion. 
In our case, we iteratively solved an optimal control system determined from a Lagrangian framework \cite{peirce:88,borzi.pra:02}. 
For \emph{desired state trapping} one maximizes the overlap of the wave function at the final time with a given desired state. 
This has been used before in theoretical works about transferring or splitting a BEC by continuously transforming a trap potential from an initial to a final shape, \emph{without} exciting the BEC \cite{hohenester.pra:07,winckel:08}. 
In the present work, we employ the same techniques, but choose the first excited state of the Gross-Pitaevskii equation at the final trap position as desired state.

For such excited state preparation, it is crucial to use an anharmonic trap, as for a weakly interacting system in a harmonic trap, displacements generate displaced ground states, i.e., coherent states. 
The present optimal control calculations have been performed in 1D (along $y$) for simplicity, using an effective 1D interaction parameter \cite{grond.pra:09b}. 
This works well, since the dynamics in the axial $x$-direction is orders of magnitudes slower than in the radial ones. 
Moreover, the potential is sufficiently anisotropic in the radial plane, such that the $z$ direction is not significantly affected by the movement, even though it is not strictly performed along $y$ (see above). 
We performed 2D simulations of the Gross-Pitaevskii equation and found only small deviations as compared to the 1D case. 

To estimate the excitation efficiency from experimental data we fit a model of the time-dependent momentum density along $y$ in the main cloud (i.e. not including the emitted atoms) after the excitation pulse (see figure 2b,c in the main text), according to
\begin{equation*}
\begin{split}
n(k_y,t) &=(1-\eta_d(t)) |\sqrt{1-\eta_e}\psi_0(k_y) + \sqrt{\eta_e}e^{i\epsilon/\hbar \cdot t} \psi_1(k_y)|^2  \\
&+ \eta_d(t) |\psi_0(k_y)|^2,
\end{split}
\end{equation*}
where the total density of the experimental data is normalized for each time step. The first line corresponds to the coherent two-level dynamics between the radially excited state $\psi_1(k_y)=\braket{k_y|1,0;0}$ and the ground state $\psi_0(k_y)=\braket{k_y|0,0;0}$ at a beating frequency given by the single-particle energy difference $\epsilon\approx E^{(1)}_y$.
The efficiency of excitation is denoted as $\eta_e$. 
In the second line, an incoherent admixture of the ground state is added with a time-dependent population $\eta_d(t)$, which corresponds to dephased atoms having undergone emission into the $\ket{0,0;\pm k_0} $ modes and subsequent collisions with the excited main cloud. 
The momentum-space wave functions $\psi_1$ and $\psi_0$ correspond to eigenstates of the one-dimensional Gross-Pitaevskii equation at a typical atom number. 
From the fit we deduce an efficiency of the coherent excitation of $\eta_e=0.97$ and a dephased fraction linearly rising from $\eta_d(\SI{5.2}{ms})=0.08$ to $\eta_d(\SI{8}{ms})=0.28$. 
For the energy difference we obtain $\epsilon=h \cdot\SI{1.78}{kHz}$.

\subsection{Emission dynamics model}

We compare the fraction of emitted high-momentum atoms to a simple model, taking into account spontaneous emission only, and assuming an atom number corresponding to the average of the shown dataset ($\bar{N}_{tot}\approx700$).
To obtain the theory curve, we solve rate equations between the states $\ket{0,0;0}$, $\ket{1,0;0}$ and $\ket{0,0;\pm k_0}$. 
The excitation rate from $\ket{0,0;0}$ to $\ket{1,0;0}$ cannot be directly obtained from GPE calculations, as both the populations of the radial states as well as their wave functions are time-dependent.
It is thus approximated by a constant rate during the excitation pulse (see ref.~\cite{Buecker2012}).
On the other hand, to treat the emission from $\ket{1,0;0}$ to $\ket{0,0;\pm k_0}$, we assume an axial density distribution according to \cite{Gerbier2004} and employ Fermi's golden rule for indistinguishable bosons to obtain a weakly atom-number-dependent two-body decay constant of the order of $\Gamma_{em}[N_{exc}]\sim 0.05\si{s^{-1}}$.
This allows to estimate the emission rate as $\dot{N}=2\Gamma_{em}[N_{exc}(t)] N_{exc}(t)^2$, where $N_{exc}$ denotes the number of atoms in $\ket{1,0;0}$, calculated as described before.
While during the initial phase, where $N_{exc}, N \ll N_{tot}$ the model works well, it fails at later times where $N_{exc}, N \lesssim N_{tot}$, indicating that stimulated processes have to be taken into account, analogous to parametric amplification in quantum optics.

\subsection{Fluorescence detection and noise corrections}

To detect the atoms after release from the trapping potentials we employ a fluorescence detector introduced in \cite{Buecker2009}. 
The expanding cloud falls through a thin horizontal light sheet (vertical waist radius $w_0=\SI{20}{\micro\metre}$) after \SI{46}{ms} of expansion time. 
Light scattered by the atoms is collected by an objective outside the vacuum vessel and imaged on an EMCCD camera. 
The extraordinarily low background of the system enables single-atom sensitivity while maintaining the dynamic range needed for imaging of dense Bose-Einstein condensates.
For the excitation power employed (which is of the order of the atomic saturation intensity at the centre of the light sheet), $\bar{p}=12.3\pm0.9$ photons per atom are detected on average. 
This number is deduced from cross-calibration by comparison of a sufficient number of calibrated absorption imaging and fluorescence shots of clouds prepared under identical conditions.
It is compatible with independent estimates obtained from physical properties of clouds released from a nearly isotropic trap~\cite{Perrin2010}.
Slight dependencies of $\bar{p}$ on the spatial position within the image due to CCD etaloning and inhomogeneity of the illumination can be corrected by using a reference image.

For exact analysis of the pair correlation as described below, also the variance of the distribution of photons per atom $\sigma^2_p$ has to be estimated.
The first contribution to this variance is the photon shot noise $\sigma^2_{SN}=\bar{p}$.
Furthermore, the excess noise due to the stochastic amplification in the EMCCD detector has to be taken into account and increases the detection noise by a factor of two \cite{Basden2003}: $\sigma^2_{amp}=2\bar{p}$, in accordance to independent characterization measurements using Poissonian light sources and the value for the detection background (see below).
Finally, a further broadening of the distribution of $p$ is caused by the diffusion of atoms within the light sheet \cite{Buecker2009}.
However, this contribution is hard to obtain from experimental results (as we cannot prepare single atoms deterministically as e.g. in \cite{Fuhrmanek2010a}) and can only be estimated from simulations.
To avoid overestimation of detection noise, which would spuriously reduce $\xi^2$, we assume $\sigma^2_p=\sigma^2_{amp}$ and set $\sigma^2_p/\bar{p}=2$ in our analysis.

Apart from the distribution of $p$, another detection contribution to $\sigma^2_s$ originates from the residual background signal $b$.
We extract $\bar{b}$ and $\sigma^2_b$ from regions directly adjacent to the main analysis regions, which contain the emitted clouds and scale the obtained values accordingly.
We obtain $\sigma^2_b/\bar{b}=2.14$, which can be understood as combined effect of shot and amplification noise and residual readout noise.
Again, to avoid over-correction, we make sure that the background treatment does not lead to spurious reduction of $\xi^2$ when the analysis regions are enlarged artificially.

\subsection{Correlation analysis}

As a first step to determine the amount of population correlation between the emitted clouds, we count the photon numbers $S_1, S_2$ (indicating atom numbers $N_1, N_2$) within regions encompassing them (see figure 3b in the main article).
Choosing the integration range along the axial $x$-direction is a sensitive task, as a too small region will not capture all emitted atoms, deteriorating the result.
On the other hand, one has to take into consideration residual overlap of the emitted clouds with the quasi-BEC.
As expected, we find that setting the limit to the position of the density minimum between main and emitted clouds minimizes the obtained value of $\xi^2$.
Still the residual overlap appears to be one of the limiting factors to the measurable correlation.
The outer limits are chosen in a way that the weak clouds emitted from $\ket{2,0;0}$ (on average less then one atom per image) are still contained within the analysis regions.
Those overlap significantly with the main emitted clouds, thus limiting the ability to assess the correlations between them separately. 

To calculate $\xi^2$ from the accessible quantities $\sigma^2_s$ and $\tilde{S}$, as defined in the text, we have to estimate the total contribution of detection noise $\sigma_d^2$ from the quantities $\sigma_b^2$ and $\sigma_p^2$ as defined in the previous section.
We can decompose $\sigma^2_s$ into a population fluctuation term and a detection term using the law of total variance:
\begin{equation*}
\begin{split}
\text{Var}(s)	=& \text{Var}(\text{E}(s|n)) + \text{E}(\text{Var}(s|n)) \\
						=& \text{Var}(\text{E}(p) n) + \text{E}(\text{Var}(p) N + \text{Var}(b)), \\
\end{split}
\end{equation*}

where $b$ and $p$ are the background signal and the number of photons per atom as discussed above, $s=S_1-S_2$, $n=N_1-N_2$, $N=N_1+N_2$. $\text{E}(X)$ and $\text{Var}(X)$ denote expectation value and variance of the random variable $X$. For our analysis, assuming $\text{Var}(X)=\sigma_X^2$ and $\text{E}(X)=\overline{X}$, this reads

\begin{equation*}
\begin{split}
\sigma_s^2		=& \bar{p}^2 \xi^2 \bar{N} + \sigma_p^2 \bar{N} + \sigma_b^2\\
						=& \bar{p}\xi^2\tilde{S} + v\tilde{S}+\sigma_b^2\\
						=& \xi^2 \sigma_{bin}^2 + \sigma_d^2,\\
\end{split}
\label{eq:tot_var}
\end{equation*}
and thus $\sigma_d^2=v\tilde{S} + \sigma_b^2$. Solving for $\xi^2$ leads to the formula used for the analysis.

To estimate the uncertainty of $\xi^2$ we propagate the standard error of the shot-to-shot mean of $b,\sigma_b^2,S$ and $p$, the latter obtained from an independent measurement as described earlier. The error of $\sigma_s^2=(k-1)^{-1}\sum_{m=1}^{k} (s_m-\bar{s})^2$, where $k$ denotes the number of experimental runs, is calculated as $\sigma_s^2\cdot (2/(k-1))^{1/2}$.

\bibliographystyle{apsrev4-1}
%\bibliographystyle{naturemag}
%\bibliography{bib_wing_paper_np}

%merlin.mbs 2010-03-15 4.21a (PWD, AO, DPC)

\end{document}